\documentclass[12pt, preprint]{emulateapj}
\usepackage{lscape}

\shortauthors{Shi et al.}

\begin{document}

\title{Morphology of $Spitzer$ 24$\mu$m-Detected Galaxies in the UDF: the Links between the Star Formation and Galaxy Morphology\\
}

\author{Y. Shi\altaffilmark{}, G. H. Rieke\altaffilmark{}, C. Papovich, P. G. P\'erez-Gonz\'alez, E. Le Floc'h }
\altaffiltext{}{Steward Observatory, University of Arizona, 933 N Cherry Ave, Tucson, AZ 85721, USA}

\begin{abstract}
We  have studied  the  morphologies of luminous infrared  galaxies (LIRGs;
$L_{IR}(8{\textendash}1000{\mu}m)>10^{11}L_{\odot}$)  at
$0.3\leq  z<1.4$ in  the $HST$  Ultradeep Field  (UDF)  by calculating
concentration  and asymmetry  indices and  comparing the  results with
similar calculations  for: (1) galaxies  at similar redshift  that are
less infrared-active; and (2) local LIRGs. We find that the
high-redshift  samples are dominated  by galaxies  with concentrations
similar  to   local  late-type   disk  galaxies;  however,   they  are
significantly more asymmetric than most local galaxies but are similar
in  both  regards  to  local  LIRGs.  On  average,  the  high-redshift
infrared-active galaxies  are  slightly more asymmetric  than the
less-active  ones, although  they  do include  a significantly  higher
portion of  highly asymmetric (merging?)  systems and  a lower portion
of   more  concentrated,   symmetric  ones.   The  morphological
similarity of  infrared-active and typical  infrared-inactive galaxies
at  high-redshift suggests  that  they  may be  from  the same  parent
population,  but are in  different stages  of an  episodic  star formation
process. The similarity between high-redshift and local LIRGs suggests
that  a  certain  level  of  asymmetry is  generally  associated  with
LIRG-level activity.

\end{abstract}


\section{Introduction}

Two  fundamental aspects of  galaxy formation  and evolution  are mass
assembly and morphology transformation.   While mass assembly is often
accompanied by strong episodes of  star formation, the morphology of a
galaxy  reflects its  dynamical  history and  evolution. For  example,
different Hubble  types are associated with  different motion patterns
of  stars and  gas and  different  histories of  star formation.   The
interplay  between these  two processes  may  play a  crucial role  in
galaxy  formation  and  evolution:  the  dynamics,  indicated  by  the
morphology, drives the star formation, while the depletion of gas from
star formation as well as feedback from young stars through supernovae
and stellar  winds may re-shape  the galaxy morphology.  Thus,  in the
local  Universe, later-type  galaxies are  gas-rich and  the  sites of
elevated  star  formation  \citep[for a  review,  see][]{Kennicutt98},
while  the  more extreme  star-forming  galaxies  often show  peculiar
morphologies,     resulting     from     collisions    and     mergers
\citep[e.g.][]{Soifer84,  Cutri85,  Sanders88,  Armus87,  Kleinmann88,
Barton00}.

The comoving  star formation  density rises by  an order  of magnitude
from  $z  =  0$  to  z~$\sim$~1  \citep{Lilly95,  Ellis96,  Steidel99,
Hopkins04}.   In the  high-redshift universe,  the high  rate  of star
formation seems to be associated with a large fraction of peculiar and
morphologically  disturbed galaxies.   Deep surveys,  especially those
carried out  by the  Hubble Space Telescope  ($HST$), reveal  that the
fraction of irregular  and interacting galaxies increases dramatically
toward high  redshift, from around  4\% \citep{Marzke98} in  the local
universe to more than 80\% at z $\geq$2 \citep{Conselice05}.  Even the
high-redshift  galaxies classified as  traditional Hubble  types still
show many  peculiar features, with  tidal or disturbed  structures and
distorted  disks. However, \citet{Melbourne05}  study  the optical  morphology
evolution  of  LIRGs, finding  that  peculiar-to-spiral LIRG  fraction
increases at lower  redshift.  Our goal in this paper  is to probe how
the  star formation  and  galaxy morphology are linked out to z$\sim$1.

The  \citet{Hubble26, Hubble36} galaxy  classification as  modified by
\citet{Sandage61},        \citet{Sandage87},        \citet{Sandage94},
\citet{Bergh60a,  Bergh60b}  and  \citet{Vaucouleurs59}, has  achieved
remarkable  success   in  categorizing  local   galaxies  with  normal
morphology  \citep[for  a  review,  see][]{Roberts94}.   However,  the
Hubble scheme has a single class for peculiar morphologies and thus is
unable to  reflect the  vast variations within  this category.   It is
therefore  inadequate to  classify the  morphologies  of high-redshift
galaxies  using  the  local  Hubble sequence,  because  a  significant
fraction of  the population are irregular galaxies  or normal galaxies
but  with perturbed  features  \citep{Driver95, Driver98,  Abraham96a,
Abraham96b,  Brinchmann98,  Cassata05}.   A  different  classification
method, called  the $CA$ system,  is particularly useful  for peculiar
galaxies at  high redshift.  It is based  on morphological parameters:
the  light  concentration   index,  $C$  \citep{Okamura84,  Abraham94,
Bershady00,  Conselice03},   and  the  asymmetry   index,  $A$,  first
introduced by \citet{Abraham94}  and developed by \citet{Conselice00}.
Its  primary virtues  are  that it  is  objective and  can be  applied
consistently to  low resolution  and low signal-to-noise  images.  The
success in  distinguishing between different Hubble  types; the robust
measurements of  parameters over  a large range  of redshift  range of
galaxy  types  especially  quantifying  the irregularity  of  peculiar
galaxies; and the strong relations between the parameters and physical
properties  such  as  color,  all  make the  $CA$  system  useful  for
classifying galaxies found in deep fields.

Out to z $\sim$ 1.5, the  star formation density is dominated by LIRGs
\citep{LeFloch05, Perez-Gonzalez05}.   Studies of the  morphologies of
these high-redshift LIRGs are necessary to understand how morphologies
and star formation are coupled  to drive galaxy evolution.  In
this    paper,   we   utilize   a   Multiband    Imaging   Photometer
\citep[MIPS;][]{Rieke04}  deep  survey of  the  $HST$ Ultradeep  Field
(UDF).  The extremely  deep $HST$ images in the UDF  allow us to apply
the $CA$  system to analyze the  optical morphologies of  the LIRGs in
the MIPS  survey  even including  faint  features  with low  optical
surface  brightness. 

The  morphologies of  galaxies in  the UDF have  been recently  studied by
\citet{Elmegreen05}  who mainly  focus on  the distribution  of galaxy
types and  \citet{Menanteau05} who  study the contribution  of various
types of  galaxies to  the cosmic star formation rate  (SFR).   The detailed analysis  of the
morphologies of  the high-redshift LIRGs and the  comparisons to other
MIPS-non-detected galaxies  and local  LIRGs in this  paper complement
these  studies and  enable us  to gain  new insight  to  the following
questions: How  do the intensely  star-forming galaxies at z  $\sim$ 1
relate to  other galaxies at the  same epoch?  How are  star formation
and galaxy morphologies coupled to drive galaxy evolution?

In this paper, we measure the concentration and asymmetry indices from
the  $HST$  images to  quantify  the  morphologies  of the  sample  of
high-redshift  LIRGs   in  the  UDF,   of  a  comparision   sample  of
MIPS-non-detected galaxies in the UDF, and of a local sample of LIRGs.
In  \S ~2,  we summarize  the observations,  data reduction,  and band
merging.  In Section  \S ~3, we describe the  selection of the samples
and  the measurement  of  the  structural parameters.   In  \S ~4,  we
present the morphological characteristics of the galaxies.  In Section
\S ~5, we discuss  the implications for high-redshift galaxy evolution
based on the results from the preceding section. In \S ~6, we present our
conclusions. Throughout  this paper, we assume  $H_{0}$=70 km s$^{-1}$
Mpc$^{-1}$,  $\Omega_{0}$=0.3 and  $\Omega_{\Lambda}$=0.7.  We use  AB
magnitudes         throughout,         where         $m_{AB}         =
23.9-2.5\log(f_\nu/\mathrm{1\,{\mu}Jy})$.  We denote  magnitudes  from the
$HST$  ACS passbands  F435W, F606W,  F775W, and  F850LP  as $B_{435}$,
$V_{606}$, $i_{775}$, and $z_{850}$, respectively.

\section{The Data}
\subsection{$Spitzer$ Data}

Our 24 $\mu$m MIPS \citep{Rieke04} observations of the UDF are part of
a scan map  covering a larger area centered on  the Chandra Deep Field
South \citep[CDFS;][]{Giacconi02}. The  images were processed with the
MIPS  instrument  team  Data Analysis  Tool  \citep[DAT;][]{Gordon05}.
Descriptions of 24 $\mu$m source detection and photometry are given by
\citet{Papovich04}.  Table  1 lists  the sources detected  at 24$\mu$m
with  unambiguous optical counterparts  that are  the primary  subject of
this  paper.  The  CDFS  was also  observed  under GTO  time with  the
Infrared Array Camera (IRAC) \citep{Fazio04}  at 3.6, 4.5, 5.8 and 8.0
$\mu$m.  Source detection  and  photometry were  done with  SExtractor
\citep{Bertin96} following similar procedures as in \citet{Huang04}.

\subsection{$HST$ Data}

The UDF is a public $HST$ survey to image a single Advanced Camera for
Surveys (ACS) wide  field camera (WFC) field (11.5  arcmin$^{2}$) in 4
broad-band      filters     (F435W;     F606W;      F775W;     F850LP)
\citep{Beckwith05}.  The  WFC  has  a resolution  of  $\sim$0.12$''$,
corresponding to  540 pc at z=0.3  and 1 kpc at  z=1.0. For our
analysis we  use the reduced  UDF data v1.0  made public by  the Space
Telescope   Science  Institute   (STScI)  on   09  March   2004.   The
$z$-band-based  catalog is  used  for optical  identification of  MIPS
sources.  A total  of 7016 objects detected at  $z$-band gives a source
density of 0.169 arcsec$^{-2}$. The 10-$\sigma$ limiting magnitudes at
$B$-band,  $V$-band, $i$-band and  $z$-band are  28.7, 29.0,  29.0 and
28.4, respectively, in an  aperture of 0.2 arcsec$^{2}$, corresponding
to 10-$\sigma$ surface brightness of 27.82, 28.12, 28.12 and 27.52
arcsec$^{-2}$, respectively.

We  also  use   Near-Infrared  Camera  and  Multi-Object  Spectrometer
(NICMOS)  UDF Treasury Observations  for the  photometry data  for our
photo-z code.   NICMOS  UDF  Observations  use  camera  3  with  a
resolution of 0.2$''$/pixel (1.6 kpc at  z=1). Due to the small field of
NICMOS  camera,  the  NICMOS   UDF  only  covers  a  subsection  (5.76
arcmin$^{2}$) of  the optical UDF.  The data reduction  and photometry
are  given by  \cite{Thompson05}. The  catalog contains  1293 objects,
giving  a  source density  of  0.0624  arcsec$^{-2}$.  The  5-$\sigma$
limiting  AB magnitude  is 27.7  at 1.1  and 1.6  $\mu$m in  a 0.6$''$
diameter  aperture, corresponding  to  a surface  brightness of  27.01
arcsec$^{-2}$.

\subsection{Results and Band Merging}

We identified 52 sources with 24 ${\mu}$m flux density greater than 0.06
mJy  ( at  which level  the  completeness is  $\sim$ 50\%  in the  UDF
\citep{Papovich04}  ).   We  searched  for  ACS  counterparts  on  the
$z$-band  image of  the UDF  within a  2.5$''$-radius  circular region
surrounding each 24$\mu$m source.  The choice of this search radius is
motivated by  the large FWHM of  the MIPS 24  $\mu$m PSF ($\sim$6$''$)
and   the  positional  accuracy   at  24   $\mu$m  (   $\sim$  0.5$''$
rms). 2.5$''$ (10 kpc at z=0.3 and 20 kpc at z=1.0) is large enough to
account  for the  possible  physical shift  between  the locations  of
infrared  emission   and  the  optical  emission   observed  in  local
interacting galaxies \citep[e.g.][]{LeFloch02}.

Multiple optical sources are present in the search circle for 41 of 52
MIPS sources.  The IRAC images  were used to identify the most probable
optical counterpart for these 41  MIPS sources.  In the UDF, there are
a  total of  312 objects  in  the IRAC  4.5 $\mu$m  catalog with  flux density
greater  than 1.6  $\mu$Jy at  50\% completeness.  23 of  the  41 MIPS
sources have  such an IRAC  counterpart within a 1.0$''$  radius.  The
probability  of random  spatial  superposition of  IRAC  sources in  a
1.0$''$-radius  circular region  is  2.3 \%,  which  indicates a  high
probability that  the IRAC counterparts are physically  related to the
MIPS sources. We  also visually inspected the IRAC  image of each source
to check if there is nearby IRAC source which might contaminate the 24
$\mu$m  emission. MIPS-15950, which  is contaminated  in this  way, is
excluded.  The  search radius to  match optical counterparts  for IRAC
sources is 1.5$''$, motivated by similar reasons to those for the MIPS
sources.  15  of 22 IRAC sources  have only one  optical source within
the  search radius,  which should  be the  optical counterpart  of the
corresponding  MIPS source.  For  the remaining  25 MIPS  sources with
possible  multiple  optical  counterparts,  if an  optical  object  is
brighter than other optical objects within the same search circle by 4
magnitudes,  this  optical object  was taken to  be the  optical
counterpart of the MIPS source.  This association is justified because
IR-luminous  galaxies  tend to  be  associated  with optically  bright
galaxies \citep{Mann97}. On the  other hand, the probability of random
spatial superposition is an order of magnitude larger for 4-magnitude
fainter objects.  Based  on the above series of  steps for deconfusion
of multiple objects  in the 24 $\mu$m beam, 16 of  52 MIPS sources may
still  have multiple  optical counterparts.  We  do not  use these  16
sources in our  analysis to avoid their affecting  the distribution of
the morphologies of the MIPS-detected sources. Our final sample therefore 
consists of 35 MIPS-detected galaxies that have unambiguous identifications with ACS images.

An   alternative  method   for   evaluating  interactions,   $''$pairs
statistics$''$, would be strongly  affected by the handling of sources
with multiple possible counterparts. However, the CA analysis employed
here measures  the structure of the  central image and  hence to first
order is  independent of  such effects. As a test, we recomputed the classification parameters including the objects
with  multiple identifications and  using the  brightest object  as the
optical counterpart.  We found that the basic  results were unchanged;
the  mean asymmetry  of the MIPS-detected  objects  got slightly
smaller but by only 0.03.

Eighteen of our sample of 35 MIPS sources have spectroscopic redshifts based
on the VLT/FORS2 spectroscopy catalog \citep{Vanzella05} and VIMOS VLT
Deep Survey  (VVDS) v1.0 catalog \citep{LeFevre04}.  For the remaining
17  sources,  we use  photometric  redshifts.  The publicly  available
COMBO-17 survey of the CDFS provides photometric redshifts by means of
17-band photometry  \citep{Wolf04}.  The typical  photometric redshift
accuracy is  ${\delta}z/(1+z)$ $<$ 0.02 for galaxies  with $m_{R}<$ 22
and  reaches 0.1  at $m_{R}\sim$24,  allowing the  rest-frame absolute
magnitude to  be accurate within 0.1  for $m_{R}<$ 22 and  0.5 mag for
$m_{R}\sim$24.   We  searched for  COMBO-17  counterparts  for each  ACS
source  within  a radius  of  0.45 arcsec.  Given  a  total of  63,501
COMBO-17  objects  in  a  field  of view  31.5$'$$\times$  30$'$,  the
probability  of  random  spatial  superposition in  a  0.45$''$-radius
circular region is  only 1.1\%.  We visually inspected  the ACS image of
each source to assure we matched the right COMBO-17 counterpart.  Seven MIPS
sources  have  X-ray  emission  based   on  the  X-ray  1  MS  catalog
\citep{Giacconi02}     and    photometric    redshifts     given    by
\citet{Zheng04}. For the ACS counterpart  of any MIPS source without redshifts from these sources,    
we    used     our    own     photo-z    code
\citep[see][]{Perez-Gonzalez05}   based  on   ACS,  NICMOS   and  IRAC
observations.  We  also used our code  to compute the  redshift for any
ACS source  with a  COMBO-17 counterpart with  $m_{R}>$ 22.   The main
advantage of  this photo-z  code is it  uses photometry over  a longer
wavelength  baseline  (including  the   1.6  $\mu$m  bump)  to  obtain
redshifts   for   galaxies   up   to  z$\sim$3,   with   accuracy   of
${\delta}z/(1+z)$  $<$ 0.1.   We inspected  the fitting  individually to
assure our code  gave a robust measurement of  redshifts.  Except for
MIPS-5106,   sources   with   multiple   redshifts   have   consistent
measurements  of  redshift.  For  this  source,  we  use our  redshift
measurement since the 1.6 $\mu$m stellar bump is obviously detected at
the IRAC bands in the SED.

The K-corrections for the optical  four-band data are computed using codes
based on  the template  spectra of galaxies  in the Sloan  Digital Sky
Survey \citep[SDSS;][]{Blanton03}.

We used the ACS image that is closest to rest-frame $B$ band to compute
the  structural parameters  for the  galaxies.  Therefore  the maximum
redshift in  our study is 1.4,  beyond which the  observed $z$-band is
shorter than the rest-frame  B-band.  The advantages of classification
using  rest-frame optical  bands and  the importance  of morphological
K-corrections  are   discussed  further  in   \citet{Windhorst02}  and
\citet{Papovich03}.

To obtain accurate total far-infrared luminosities, it is necessary to
have photometry over the whole range of infrared wavelengths.  In this
study, however  it is enough  to have a  coarse estimate of  the total
infrared  luminosity.   We  used  the  template  spectra  of  starburst
galaxies (with the current SFR much larger than the past-averaged SFR) published by \citet{Lagache04} to get the 8-1000 $\mu$m total
infrared luminosity from the observed 24 $\mu$m flux density. Due to the high
sensitivity of  MIPS, some MIPS-detected galaxies at  low redshift are
normal  galaxies, not  starbursts. For  them, we  used  the appropriate
template for  galaxies with total  IR emission fainter  than 10$^{10}$
L$_{\odot}$  (where  the  threshold  is computed  from  the  starburst
template).   The  average  conversion  factor  from  ${\nu}L_{\nu}$(24
$\mu$m) to total infrared luminosity is  20 with a dispersion of up to
0.2 dex,  indicating that our conversion should  be accurate typically
within a factor 2 or  3 \citep{Papovich02, LeFloch05}, which is enough
for the analysis in this paper.

\section{Methodology}

\subsection{Sample Definition}
We only computed the structural parameters for 34 of the 35 MIPS sources
that have signal-to-noise (S/N) ratios better than 100 and half
light radius $r_{50}$ $>$ 5  pixels at $z$ band.  High signal-to-noise
ratio  (SNR)  and large  size  are  required  for robust  measures  of the
asymmetry parameters \citep{Conselice00, Bershady00}. The  deep exposure in the UDF thus
allowed  us  to  compute  reliably  structural parameters  for  a  high
fraction (97\%) of  MIPS-detected galaxies.

The members of  our control sample are defined to  be undetected at 24
$\mu$m and (1) to  lie 5 $''$ away from all MIPS  sources; (2) to have
half light radius  r$_{50}$ $>$ 5 pixels at $z$ band;  (3) to have SNR
better than  100 at $z$ band;  and (4) to have  a COMBO-17 photometric
redshift  smaller  than  1.4.  The  last condition  is  necessary  for
K-corrections  and  requires  $m_{z}<$ 25  \citep{Wolf04}.   The  final
sample is composed of  252 normal galaxies, representing 80\% of
the  galaxies undetected  at  24$\mu$m meeting  only  the first  three
requirements.

Our  main goal  is to  investigate the  morphologies  of high-redshift
LIRGs. We identified a galaxy with $L_{IR}$ $>$ 10$^{11} $L$_{\odot}$ as
a LIRG; the final sample of LIRGs has 21 sources. The deep exposure of
the UDF  allows us  to detect their  optical counterparts down  to low
surface   brightness,  removing   potential   biases  from   less-deep
classification imagery. The 10-$\sigma$ surface brightness at $z$-band
is  27.52  arcsec$^{-2}$, which  corresponds  to  a rest-frame  B-band
surface  brightness  of 24.5  for  a  fiducial  galaxy at  z=1.   This
corresponds well  to the typical  surface brightness level  reached in
morphological studies  of local  galaxies (although at  lower physical
resolution). It is  also close to the central  surface brightness of a
local  low-surface-brightness  (LSB) galaxy.   In  addition, the  LIRG
sample is sufficiently  large for reliable statistics, as  can be seen
from the results in Table 2.

We  also created  a  sample  of 5  MIPS-detected  but lower  luminosity
non-LIRGs and a control sample of 137 MIPS-non-detected galaxies.  The
sample  of  MIPS-detected  non-LIRGs  is  composed  of  galaxies  with
$L_{IR}$ $<$ 10$^{11}$ L$_{\odot}$  and $M_{B}$ $< -18.5$.  The sample
of MIPS-non-detected galaxies is derived from the sample of 252 normal
galaxies with  the additional constraints  that $M_{B}$ $<  -18.5$ and
that the redshift  be greater than the minimum  redshift of the LIRGs.
The cut-off magnitude $M_{B}  = -18.5$ is approximately the rest-frame
$B$ magnitude of the faintest LIRG.

\begin{figure}
\epsscale{1.0}
\plotone{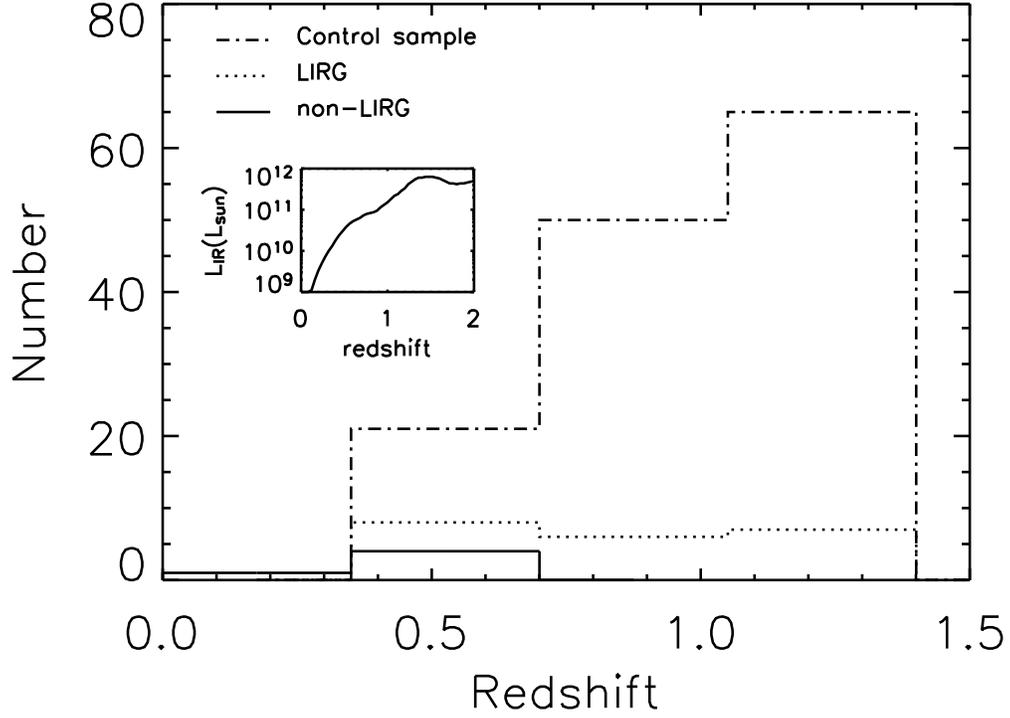}
\caption{\label{Redshift} The   redshift   distributions    of   LIRG ($L_{IR}$(8-1000$\mu$m)$>10^{11}$L$_{\odot}$),
non-LIRG ($L_{IR}$(8-1000$\mu$m)$<10^{11}$L$_{\odot}$)   and
MIPS-non-detected control samples. The  dot-dashed histogram  indicates the
control sample, the dotted histogram indicates the LIRGs and the solid
histogram indicates  the non-LIRGs. The control and  LIRG samples have
almost the same redshift  distributions, while non-LIRGs are mainly
at low redshift due to the detection limit. The inserted plot shows the IR luminosity at the 24 $\mu$m detection limit (0.06 mJy ) as a function
of the redshift. }
\end{figure}

Figure~\ref{Redshift}  shows the redshift  distributions of  the three
samples and the infrared luminosity at 24 $\mu$m flux density of 0.06 mJy as a
function of  redshift.  Galaxies of  all three samples mainly  fall at
z$>$0.3.   The sample  of  LIRGs and  MIPS-non-detected galaxies  have
almost the same redshift  distributions while non-LIRGs mainly fall at
lower  redshift  due  to  the  detection  limits.   In  the  following
analysis,  we  mainly  discuss  the  morphologies  of  the  LIRGs  and
comparisons to the MIPS-non-detected galaxies, since the MIPS-detected
non-LIRGs have a different redshift range and are few in number.

\subsection{Quantitative Morphology Classification: Concentration and Asymmetry}

Before  measuring  the  concentration  and asymmetry  value  for  each
galaxy,  some image processing was  required.   We  identified and masked
contaminating  sources  in  the  region  of the  galaxy  using  $z$-band
segmentation maps provided  with the UDF data release. We replaced  the mask region
with noise to  compute the morphology parameters.

We measured the concentration  index using the methodology described in
\citet{Bershady00}.   To  study the morphologies in a fixed-size aperture at the physical scale of each galaxy,
 the  dimensionless parameter $\eta$ as defined  by $\eta(r) =
I(r)/<I(r)>$ \citep{Petrosian76} is used to define the
total size of  the galaxy, where $I(r)$ is  the surface brightness
at radius  $r$ and $<I(r)>$ is the  mean surface brightness within
radius  r. The apparent  total magnitudes are then integrated  over an
aperture equivalent to the
total size of the galaxy  defined as twice the radius $r(\eta=0.2)$.
Based  on the  measured curve  of growth,  the concentration  index is
defined by
\begin{equation}
C=5\log{\frac{r_{80}}{r_{20}}},
\end{equation}
where $r_{80}$ and $r_{20}$ are the radii that enclose 80\% and 20\%
of the total light \citep{Kent85}, respectively.

The full description of the algorithm to compute the asymmetry index is in \citet{Conselice00}. The asymmetry parameter
is defined as:
\begin{equation}
A=\min(\frac{\sum|I_{0}-I_{180}|}{\sum|I_{0}|})-\min(\frac{\Sigma|B_{0}-B_{180}|}{\sum|I_{0}|}),
\end{equation}
where the sums are over the total size of the galaxy as defined above.
In the above equation, the  first term represents the asymmetry of the
galaxy  and  the  second one  corrects  the  effect  of noise  on  the
asymmetry measurement,  where $I_{0}$ and $I_{180}$  are the intensity
of each pixel of the image  and of the image rotated by 180$^{\circ}$,
respectively,  and $B_{0}$  and $B_{180}$  are the  intensity  of each
pixel   of  the  background   region  and   its  180$^{\circ}$-rotated
counterpart,   respectively.   We   extracted  a   100$\times$100  pixel
background region without any contamination and scaled the asymmetry of
this  region by the  relative size  of the  galaxy to  this background
region  area.  The  asymmetry  minimum was  obtained  by searching  the
minimum of  asymmetries at rotation centers around  the initial galaxy
center.    See  \citet{Conselice00}   for  the   detailed  computation
algorithm  to  locate  the  rotation center  producing  the  asymmetry
minimum.

The  uncertainty  in the concentration  measurement  is  mainly due to  the
spatial resolution.   \citet{Bershady00} show that  the uncertainty of this parameter
is around  0.2 for $r_{50}>$5 pixel.  For  asymmetry measurements, the
noise of the image is  the main effect.  \citet{Conselice00} show that
asymmetry measurements have errors  around 0.02 for S/N$>$500 and around
0.05  for   100$<$S/N$<$300.   We  tested  our   programs  by  computing
the concentration   and   asymmetry   parameters   of   20   galaxies   in
\citet{Conselice03} with asymmetry spanning from 0.01 to  0.4 and concentration from 2.5 to 4.5. The
average  differences  in concentration  and  asymmetry parameters  are
 0.2 and  0.02, respectively.  For galaxies with  SNR better than 100
in  our sample,  we estimate  the typical  errors of the  concentration and
asymmetry  indices are around  0.3 and  0.04, respectively,  which are
adequate for our study.

\section{RESULTS}
\subsection{Morphologies of a Representative Sample of LIRGs at 0.3$<$z$<$1.4}

An  important  aspect   of  this  study  is  that   we  can  determine
morphological properties for a  representative sample of high-z LIRGs.
In  the UDF,  23 of  34 MIPS-detected  galaxies at  0.3$<$z$<$1.4 with
$M_{B}< -18.5$,  $S/N>100$ and $r_{50}>5.0$ are  LIRGs.  Compared with
137 MIPS-non-detected galaxies with  the same constraints as above, we
find that  $\sim$15\% of the  galaxies with $M_{B}<  -18.5$, $S/N>100$
and  $r_{50}>5.0$  are  LIRGs  at  z$\sim$1.   This  fraction  may  be
underestimated    due   to   the    non-detection   of    LIRGs   near
$L_{IR}\sim10^{11}$  L$_{\odot}$ and  also may  be overestimated  as a
result   of   missing   non-LIRGs   near  $M_{B}=   -18.5$   at   high
redshift. However,  the portion of LIRGs  in our sample  is similar to
those found by others at  similar redshift.  For example, the fraction
of LIRGs  at z$>$0.4  in the Canada  France Redshift Survey  (CFRS) is
about 16\%  without correction for the non-detection  of galaxies near
the  cut-off limits  \citep{Hammer05}.   Another test  is  to use  the
infrared and optical luminosity functions.  \citet{LeFloch05} recently
obtained the  infrared luminosity  function at 0$<$z$<$1.2  based on
the  MIPS/$Spitzer$ 24  $\mu$m  sources located  in  the CDFS.   Using
$L_{IR}^{\star}=1.9\times10^{10}L_{\odot}$,   $\alpha_{IR}=1.23$   and
$\sigma_{IR}=0.72$ at 1.0$<$z$<$1.2,  we estimate  70\% of the LIRGs at  0.9$<$z$<$1.4 are
below 0.06 mJy.  Using  the rest-frame $B$-band luminosity function of
the COMBO-17  survey \citep{Wolf03}, we estimate 65\%  of the galaxies
with $M_{B}< -18.5$ are  below $m_{B}=25$.  Correcting the fraction of
non-detected galaxies at  high redshift, we find that  the fraction of
LIRGs is almost  the same ($\sim$18\%) as in our study.  At z$<$ 0.9 where both LIRGs
and  $M_{B}<  -18.5$  galaxies  are  detected, we  obtain  a  fraction
$\sim$15\% of LIRGs for a total  of 67 galaxies.  We conclude from all
these  arguments that  the fraction  of  LIRGs for  $M_{B}< -18.5$  is
$\sim$ 15\% at 0.3$<$z$<$1.4, the same as in our study. This fraction 
is much larger than that in the local universe, which is 0.5\% for galaxies with 
$L_{\rm tot}>$ 10$^{10}$ L$_{\odot}$ \citep{Soifer86}.

Table  1  lists  the  LIRGs  detected  in the  UDF  along  with  their
concentration  and  asymmetry indices,  $C$  and  $A$. MIPS-4641, MIPS-5117  and
MIPS-15236 have X-ray detections \citep{Giacconi02} and are classified
as type 1  AGN by \citet{Zheng04}.  These three  sources are excluded in
our study of morphologies. The other two MIPS sources detected in the X-ray
are classified as galaxies by \citet{Zheng04} and are kept in this study.

Table 2  lists the  statistical results for  LIRGs, non-LIRGs  and the
MIPS-non-detected sample in the intermediate redshift bin 0.3$<$z$<$0.9, in
the  high redshift  bin  0.9$<$z$<$1.4,  and for  the  local (low redshift) LIRGs.  At
0.3$<$z$<$0.9, the 24 $\mu$m detection limit (0.06 mJy) and the B-band
magnitude  limit  (-18.5)  yield   complete  samples  of  LIRGs  and
MIPS-non-detected  galaxies,  respectively.   At  0.9$<$z$<$1.4,  both
the LIRGs and  MIPS-non-detected samples are incomplete.  We mainly discuss
our   result   based   on    the   complete   sample.   However,   the
MIPS-non-detected  sample  at  0.9$<$z$<$1.4  is still  dominated  by
non-LIRGs.   Given 9 LIRGs,  88 MIPS-non-detected  objects and  70\%
incompleteness in IR-detection at 0.9$<$z$<$1.4, the fraction of LIRGs
in 88 MIPS-non-detected objects is $<$25\%. Therefore, the comparision
of morphologies  between LIRGs and MIPS-non-detected  galaxies at this
redshift range is still helpful in our  understanding the difference
in morphologies between infrared-active galaxies and infrared-inactive
galaxies.

\begin{figure}
\epsscale{.9} 
\plotone{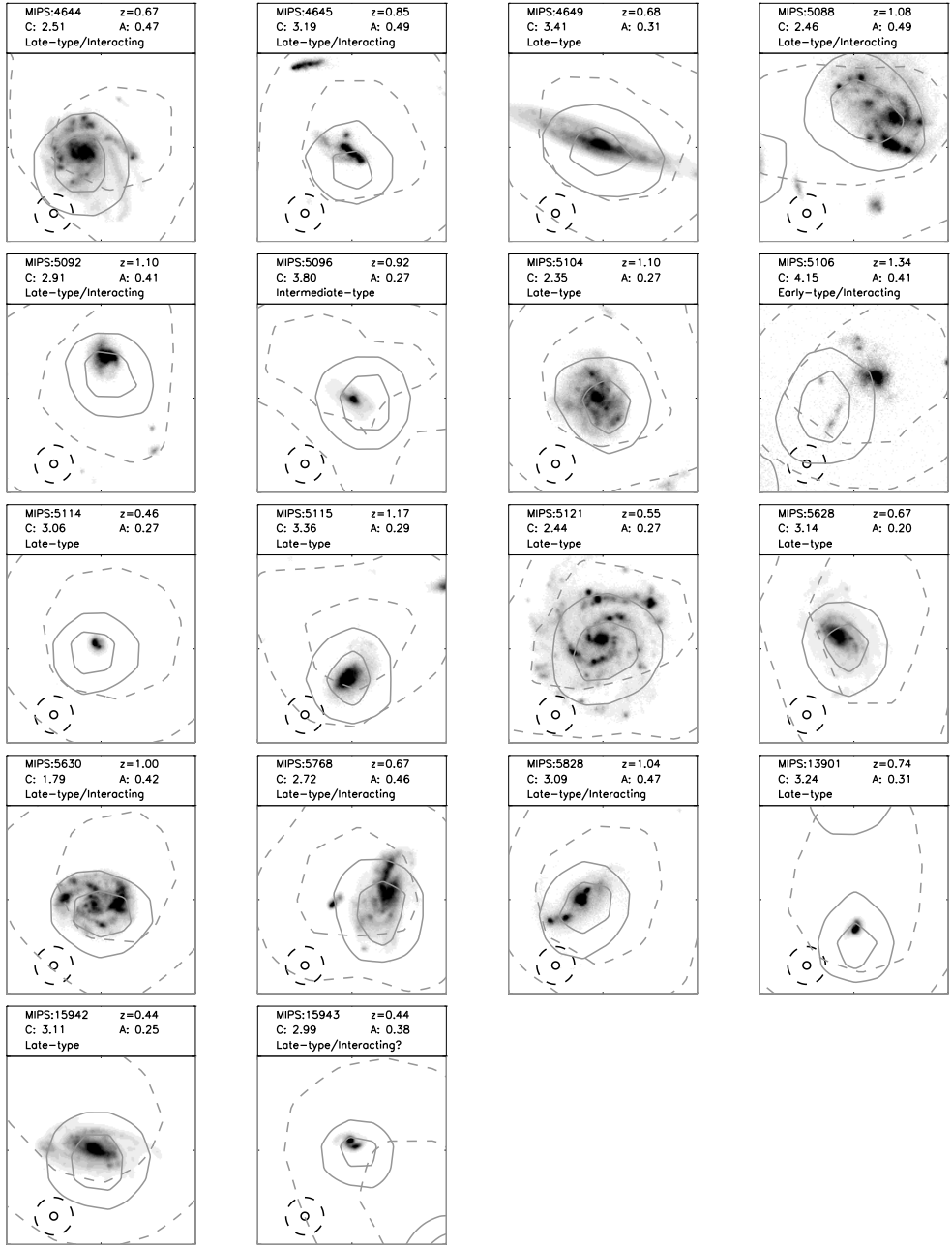}
\caption{\label{Image_LIRG} The  rest-frame B-band images (5$\times$5 arcsec)  of  non-X-ray LIRGs  in the  UDF  along with  the
redshifts,  structural parameters and  the galaxy  types based  on the
$CA$ classification scheme. The center of the image is the position of the MIPS source. 
The light solid and dashed curves show  contours at 20\% and 50\% 
enclosing flux for IRAC and MIPS, respectively. The heavy solid and dashed circles show the position accuracy for
IRAC sources (0.1 arcsec) and MIPS sources (0.5 arcsec).  }
\end{figure}

\begin{figure}
\epsscale{1.0}
\plotone{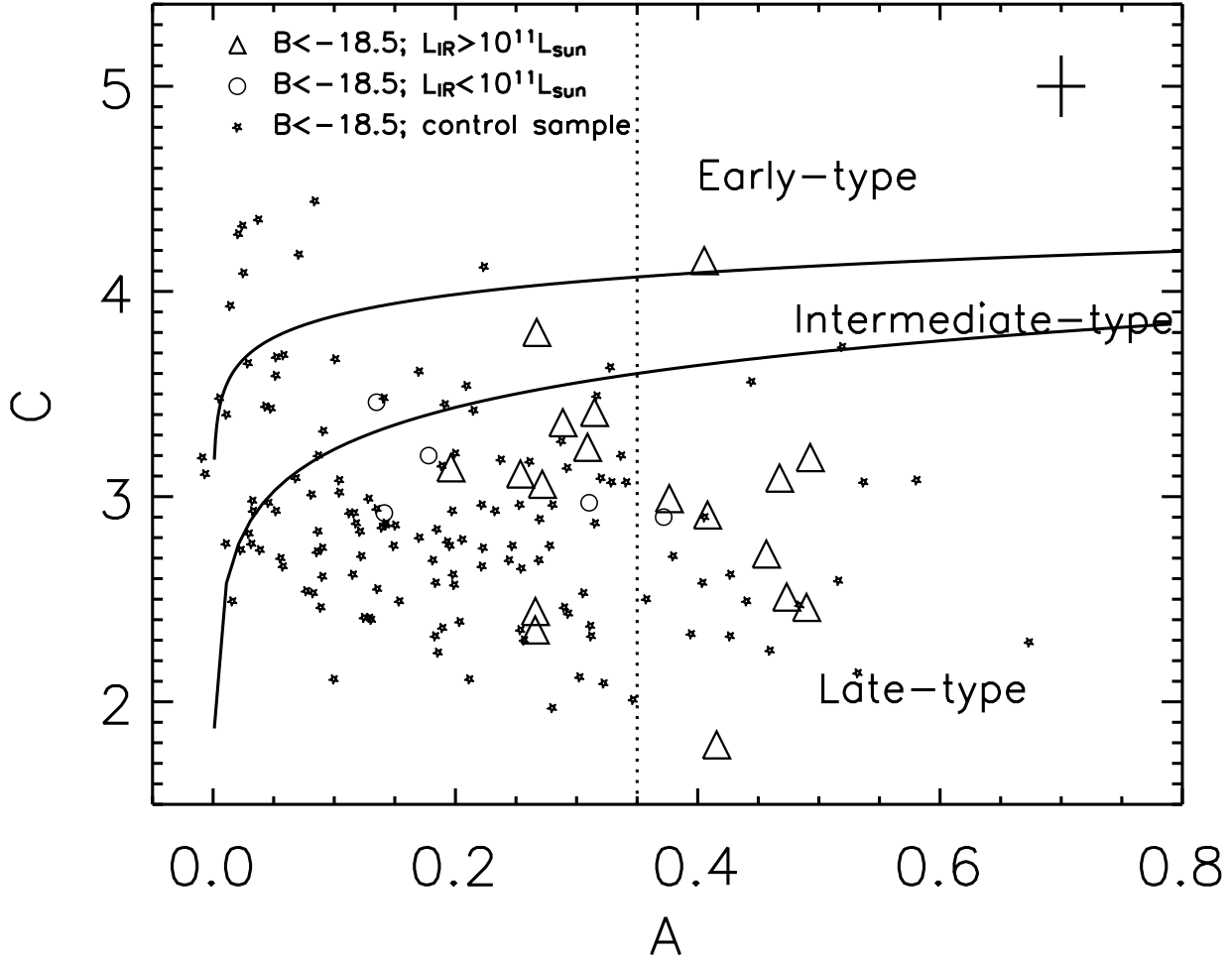}
\caption{\label{Con_asym} 
The distributions  of LIRGs (triangles), non-LIRGs  (open circles) and
MIPS-non-detected control  sample galaxies  (stars) in  the plane  of  concentration and
asymmetry.  The two solid curves divide the plane  into three regions
populated by  early-, intermediate- and late-type  galaxies. The dotted
line indicates $A=0.35$, to the right of which galaxies are identified
as merging systems. The cross shows typical errors of concentration (0.3) and asymmetry (0.04).
}
\end{figure}

Figure~\ref{Image_LIRG} shows the rest-frame $B$-band images for LIRGs
used  for  our morphological  study.  Figure~\ref{Con_asym} plots  the
galaxies  along  with our  other  two high-z  samples  on  the $C,  A$
plane. The  solid curves  show the division  of this plane  into three
regions   mainly  populated   by  early-type,   intermediate-type  and
late-type  galaxies,  based   on  classifications  of  local  galaxies
\citep[c.f.][]{Abraham96a,          Abraham96b,          Conselice00}.
Figure~\ref{Image_LIRG}  shows   that  the  high-redshift   LIRGs  are
predominantly   late-type    galaxies   \citep[as   found   previously
by][]{Bell05a} with asymmetric  structures, including long tidal tails
and distorted disks.   Some of them have very  low surface brightness,
such   as  MIPS-5106   with  rest-frame   B-band   surface  brightness
$\mu_{B0}=22.4$ within the Petrosian radius.

We define a galaxy with A$>$0.2 as an asymmetric galaxy in this study.
This is motivated  by the fact that most  ($\sim$90\%) of local normal
galaxies  including   elliptical  galaxies,  early-type   (Sa-Sb)  and
late-type   spiral   galaxies    (Sc-Sd),   have   asymmetry   A$<$0.2
\citep{Conselice03}.  As listed in  Table 2, 90$^{+10}_{-30}$\% of the LIRGs
in the  UDF are  late-type galaxies and  80$\pm$10\% of the  LIRGs are
asymmetric (A$>$0.2) objects.

The  classification of  the high-z  galaxies using  the  asymmetry and
concentration  indices is generally  consistent with  the conventional
visual classifications.  Figure~\ref{Image_LIRG} shows that only two
sources, MIPS-5115 and MIPS-13901  are classified visually
differently from the results using the structural parameters.  In this
paper, we make no use  of the conventional classifications (other than
to place  some results in context),  basing all of  our conclusions on
the  $CA$   analysis.  Thus,   the  small  number   of  classification
discrepancies has no effect on our conclusions.

\subsection{Comparision of LIRG Morphologies to those of MIPS-non-detected Galaxies}

In this  Section, we  compare  the morphologies of  LIRGs in the
UDF to  those of MIPS-non-detected  galaxies.  Table 2 shows  that the
correlations at high redshift between infrared luminosity  and morphology are weak.  On
average, high-$z$ LIRGs are more asymmetric than the MIPS-non-detected galaxies,
characterized by  a higher fraction  of $A > 0.35$  \citep[merging;][]{Conselice03} galaxies
and fewer symmetric objects, defined  as A$<$0.2. However, a number of
MIPS-non-detected  galaxies  show  asymmetric  structures  and/or  are
currently involved  in merging activity.  In the  following we compare
the morphological distributions in  terms of the fraction of late-type
galaxies and the distribution of asymmetry parameters.

\subsubsection{Distributions of the Asymmetry and Concentration Parameters}

Table 2 lists  the average asymmetries and the  standard deviations of
the means.  The LIRGs have asymmetry around 0.34, somewhat higher than
$A$=0.22  of the  MIPS-non-detected galaxies.  The K-S  test  shows the
probabilities that  LIRGs and  MIPS-non-detected galaxies have  the same
distribution  of asymmetries  are  1\% and  $<$0.1\%  for the intermediate-redshift  and
high-redshift  bins,  respectively.  This  seems  consistent  with  the
suggestion   \citep[e.g.][]{Larson78,    Sanders96,   Barton00}   that
non-symmetric structures,  such as tidal  distortions, companions, and
distorted disks, play an important role in enhancing star formation in
galaxies.  Figure~\ref{Con_asym}  indicates that the MIPS-non-detected
galaxies  have a  wide  distribution of  morphologies, from  classical
Hubble types  to highly asymmetric, most likely  merging systems.  The
fraction  of  symmetric (A$<$0.2)  galaxies  in  this  sample is  high
($\sim$  50 \%).  In  comparison, the  LIRGs virtually  lack symmetric
objects  ($\sim$10\%).  

As listed  in Table  2, the  sample of LIRGs  shows a  slightly higher
fraction of  late-type galaxies  than the sample  of MIPS-non-detected
galaxies, although the values are consistent within statistical errors.  
The classification calibration method in this comparison is
based  on  visually  classified  local  galaxies  \citep{Conselice00}.
Alternatively,   \citet{Cassata05}   measured   the   rest-frame   $B$
morphology for the K-band selected high-redshift galaxies and used the
boundary  ($A<$0.2   and  $C>$2.9)  to   define  the  early-type
galaxies.  Based on their method,  we find that the late-type fraction
of  LIRGs  is  still  a  little higher  at  0.3$<$z$<$1.5,  i.e.,  the
24$\mu$m-selected sample is biased a little toward late-type galaxies.
However, much of this relatively  minor difference can be explained by
the larger portion of  symmetric early- and intermediate-type galaxies
in  the   MIPS-non-detected  sample.   The   differences in morphologies between  the
late-type galaxies of these two samples are therefore minor.

On the  other hand, the galaxies  of both samples  are more asymmetric
than the local Hubble-type  galaxies and even slightly more asymmetric
than  the non-starburst  local  dwarf irregular  galaxies, which  have
asymmetry of  0.17$\pm$0.10 \citep{Conselice03}.  The  K-S test shows
that the probability that LIRGs  and local irregular galaxies have the
same distribution of asymmetry is $<$0.1\%, while such a probability for
MIPS-non-detected  galaxies  at   0.3$<$z$<$0.9  and  local  irregular
galaxies   is  27\%.   \citet{Conselice00}   show  that   low  spatial
resolution  affects the  asymmetry  measurement by  making the  galaxy
appear more symmetric. The  spatial resolutions are 50 pc/pixel for the
local sample of \citet{Conselice03} and 300 pc/pixel for our galaxies
at $z$=0.5.  This reinforces our  conclusion that the galaxies  in the
UDF are more asymmetric than the local galaxies.

Although  the two  high-z samples  show somewhat  different  levels of
asymmetry, their concentrations are  almost the same, characterized by
values  that  are  comparable  to the  local  late-type  disk-galaxies
\citep[Sc-Sd;][]{Conselice03}.  The  K-S test shows  that the LIRG sample
and  the MIPS-non-detected   sample  have   the   same  distribution   of
concentration  with a probability  of 40\% and 30\% in the intermediate and high
redshift  bins, respectively.   \citet{Conselice03} argues  that  the
concentration  index  reveals the  past  star-formation  history of  a
galaxy because of  the correlation of bulge to  total light ratios and
stellar  masses  with this  index.   This  behavior  implies that  the
  MIPS-non-detected   galaxies  have   similar  stellar
populations and distributions as the LIRGs except for the LIRGs having
a large  population of massive  young stars due  to the presence  of a
starburst.

\subsubsection{Merging Galaxies}

It is  also of interest to  determine how many LIRGs  are merging.  We
classify an object  as a merger if its  asymmetry parameter is greater
than 0.35  \citep{Conselice03}.  The merger  as defined here  means a
major merger,  i.e., that the two progenitor  galaxies have comparable
masses. However, other strong asymmetries may also satisfy this numerical definition. 
By this  definition, the LIRG sample is  composed of 40$\pm$20
\%  merging systems, higher  than the  merging fraction
($\sim$15\%)  of the MIPS-non-detected  sample.  Although  the merging
fraction obtained by the asymmetry  technique depends on the $A$ value
used for  merger identification, this conclusion does  not change with
modest changes in this defining value of $A$.

Our sample shows 40\% of LIRGs at 0.3$<$z$<$1.4 are merging systems as
identified  by  the  $CA$  method  (which emphasizes  high  levels  of
morphological    disturbance   in    systems    with   high    surface
brightness). This  result is probably consistent with  but higher than
that  of  \citet{Bell05a} who  found  that  less  than 30\%  of  1500
star-forming  galaxies   at  redshift  0.65$<$z$<$0.75   are  strongly
interacting   based   on   visual   classifications,   and   that   of
\citet{Zheng04}  who   found  that   6  ($\sim$17\%)  of   36  distant
(z$\gtrsim$0.4)  LIRGs   in  the   CFRS  are  obvious   ongoing  major
mergers. \citet{Zheng04}  do not count the  irregular galaxies (22\%),
which may have A$>$0.35, as  merging galaxies. Another cause for these
differences is that the morphologies  of a merging galaxy at different
surface brightness levels show  different levels of asymmetry and thus
visual identification on a relatively  shallow image may mis-classify it
as a non-merging galaxy.

\section{Discussion}

\subsection{Continuous Morphological Transformation plus Episodic Starbursts}  

Our  study of  morphologies and  infrared properties  of high-redshift
galaxies in the UDF has two basic results. The first one is that LIRGs
and MIPS-non-detected galaxies in  the UDF are predominantly late-type
asymmetric galaxies.   The LIRGs are somewhat  more asymmetric on
average than MIPS-non-detected galaxies and they are characterized by
a  higher fraction of  very asymmetric  ($A \ge  0.35$) systems  and a
lower  fraction of  symmetric ($A  <  0.2$) objects.   This result  is
consistent  with the  proposal that  non-symmetric structures  play an
important role in  enhancing the star formation and  hence elevate the
infrared luminosities of galaxies. The  second result is that there is
no  one-to-one correlation  between  the occurrence  of high  infrared
luminosities and  disturbed morphologies: highly  asymmetric galaxies,
including  merging  systems, do  not  necessarily  have high  infrared
luminosities.

On average, high-z LIRGs and  normal galaxies are both more asymmetric
than  local  irregular  types,  and their  concentration  indices  are
similar  to the  local late-type  disk galaxies.   Are  the asymmetric
normal galaxies  and the LIRGs  two different populations or  are they
the  same  population but  in  different  stages?   That is,  are  the
activities of starbursts episodic so  that the LIRGs are now in active
stages  and asymmetric  non-LIRGs  are similar  galaxies  that are  in
quiescent states between-starburst stages?

Based on  the statistics at 0.3$<$z$<$0.9  listed in table  2, 25\% of
asymmetric  galaxies  (A$>$0.2)  are  LIRGs  at  0.3$<$z$<$0.9.   This
fraction  is $\sim$10\%  in  the local  universe,  assuming a  merging
fraction   of  4\%   \citep{Marzke98},  a   LIRG  fraction   of  0.5\%
\citep{Soifer86} and  an asymmetric galaxy  fraction in LIRGs  of 74\%
based on the fraction of  close pair and strongly interacting galaxies
\citep{Sanders88}.  One  candidate population for  asymmetric galaxies
without starbursts  is interacting  galaxies with low  metallicity and
hence low dust.  However,  \citet{Kobulnicky04} find that the shift of
the luminosity-metallicity relation is smaller than 27\% at z$\sim$1.0
compared to  the local trend, based  on the nebular  emission lines of
galaxies   ($18.5<M_{B}<-21.5$)  with   SFR   $\sim$0.1-10  $M_{\odot}
yr^{-1}$.    This   result  is   consistent   with  previous   studies
\citep{Kobulnicky99,  Carollo01}.   We conclude  that  both LIRGs  and
MIPS-non-detected galaxies are not likely to be extremely deficient of
dust.   Another  candidate   population  for  asymmetric  non-detected
galaxies  is  interacting  galaxies  without gas,  e.g.,  interactions
between galaxies  composed entirely  of stars.  However,  such $''$dry
mergers$''$  appear   in  general  to   be  uncommon  ($\sim$   1  \%)
\citep{Bell05b}.     Therefore,   the    similar    distributions   of
concentration  index   for  LIRGs  and   asymmetric  non-MIPS-detected
galaxies  suggests  that  these  two samples  have  similar  formation
histories. The  moderately higher asymmetry for the  LIRGs could arise
from luminous off-nuclear star formation, or from extinction. Thus, the
elevated   periods   of   star   formation  are   probably   occurring
episodically.

In  the local  universe, starbursts   in a  given galaxy  have  a short
duration    of   0.01   to    0.07   Gyr    \citep{Krabbe94,   Lutz96,
Alonso-Herrero00}.   Near-IR  imaging  and spectroscopy in  the  starburst
galaxies  M82,  IC 342  and  NGC  253  indicate several  episodes  of
enhanced   star   formation   \citep{Rieke93,   Satyapal97,   Boker97,
EngelBracht98,  Forster-Schreiber00}.    The  anticorrelation  between
equivalent width  of H${\alpha}$  and galaxy pair  spatial separation
found by  \citet{Barton00} indicates  that the starburst episodes
are initiated  by a  series of close  passes.  In the  early Universe,
Lyman  Break Galaxies  (LBGs) show  comparable durations  of starburst
activity \citep{Papovich01,  Shapley05}.  Recently, \citet{Yan04} and \citet{Papovich05}
found  that the broadband photometry data for massive galaxies at z$\sim$1-3 are better
fit by a model star-formation history with recent bursts superimposed on an underlying
stellar population compared to simple, monotonically-evolving stellar populations. 
They  conclude  that  these  galaxies  have  a
complicated   history   with   many   episodes  of   star   formation.
Theoretically, simulations  show that in merging  galaxies, strong gas
inflow is triggered  and nuclear star formation is  ignited only after
enough   gas  is   accumulated   in  the   center   of  the   galaxies
\citep[e.g.][]{Noguchi91, Mihos93, Barnes96}. During the whole merging
process, strong enhancement of the  star formation only occurs with a
short duration, which supports the proposal for episodic starbursts. A
star formation history involving  episodic events is also indicated by
recent semi-analytic studies \citep[e.g.][]{Somerville01, Nagamine05}.

Given  the merging  timescale  $t_{\rm merge}$,  the  fraction $f$  of
$t_{\rm merge}$ at which galaxies are actually involved in starbursts,
and  the  timescale for  episodic  starbursts  $t_{\rm  ESB}$, we  can
estimate  the total  number  of episodes  during  one merging  event
$N_{\rm ESB}$=$f \times t_{\rm  merge}/t_{\rm ESB}$.  We adopt 0.5 Gyr
for    the    timescale of the greatest merger activity \citep[e.g.][]{Patton00}.     The
characteristic timescale  of starbursts is estimated  at 10-70 Myr
as shown  above.  The  fraction $f$ can  be computed from  the merging
fractions of LIRGs and infrared-inactive galaxies as listed in Table 1
and the  fraction of  LIRGs in a  given magnitude-limited  sample.  We
adopt 15\% as  the fraction of LIRGs as shown in  Section 4.1 and find
$f=0.15\times0.4/(0.15\times0.4+0.85\times0.16)=0.3$.   Therefore,  the average
number of  episodes during a merging  event is around  4 for $t_{\rm
ESB}=0.04$  Gyr.  This indicates  that during  each merging  event, at
least  one episode  of vigorous  star formation  can be  triggered and
multiple episodes of such activity are possible.

This conclusion  is supported by  the studies  of nearby
galaxies cited above. In addition, simulations  by \citet{Mihos96} show that during
merging events,  bulgeless disc galaxies  can trigger gas  inflow even
before the occurence of the  actual collision of the galaxies and thus
two  starburst episodes  can  occur during  the  merging  event.
Recently, \citet{Tissera02}  has shown that  in hierarchial clustering
scenarios, some  merging events can  have two episodes  of starbursts:
one owing  to the gas inflow  driven as the  satellite approaches, and
the second one occuring in the collisions of baryonic clumps.

\subsection{Comparisions to the Local LIRGs}

In the last column of Table  2, we present the statistical results for
local LIRGs listed in  the IRAS Bright Galaxy Sample \citep{Soifer87}.
We extracted the  image for each source from  the Digitized Sky Survey
(DSS)  at  the  POSS2/UKSTU  blue  band.  The  table  shows  that  the
morphologies   of  local   LIRGs   are  similar   to   those  of   the
MIPS-non-detected  sample and the  LIRGs in  the UDF  in terms  of the
fraction of  late-type galaxies, of merging galaxies  and of symmetric
galaxies.  The  mean asymmetry and  concentration of LIRGs in the UDF are
slightly higher  than those of local  LIRGs.  The K-S  test shows that
the probability that  the LIRGs at 0.3$<$z$<$0.9 and  local LIRGs have
the same distribution of asymmetry  and concentration are 3\% and 9\%,
respectively. That is, the apparent differences are only marginally 
significant. In the local universe, a very low (4\%) fraction of the
total    population   is    irregular    and   interacting    galaxies
\citep{Marzke98}. As a result, the morphologies of local LIRGs are
quite  different  from the  local  normal infrared-inactive  galaxies,
which is not  the case at high-redshift.  The  small difference in the
morphologies between LIRGs and infrared-inactive galaxies in  the UDF is due to the fact that,
at high  redshift, a significant  fraction of galaxies  are asymmetric
systems in  which starbursts are much more  easily triggered. Locally,
LIRGs are  exceptional, and an exceptional disturbance  is required to
trigger their elevated rates of  star formation.  Based on this trend,
we can expect  that at higher redshift (z$>$1.5),  the morphologies of
LIRGs and non-LIRGs will be even more indistinguishable.

The  similarity between the morphologies of local and high-z LIRGs
suggests that a certain level of asymmetry is correlated with vigorous
star formation, independent of redshift. This situation probably arises
because once star formation is initiated, it evolves within a
galaxy independent of the galaxy environment \citep{Balogh04}. Therefore,
a similar level of perturbation leads to similar consequences in terms
of elevated star formation. At z~$\sim$~1, most
galaxies appear to have an appropriate level of asymmetry potentially to
be transformed into LIRGs by relatively minor mergers.
Therefore, the incidence of LIRGs is high, leading to a high density of
star formation. Locally, most galaxies are highly symmetric and only
exceptional circumstances, such as an interaction with another massive galaxy,
can produce the situations that are a prerequisite for a LIRG. Hence, the incidence of LIRGs is low and
so is the density of star formation.

\section{Conclusions}

In  this   paper,  we  have   investigated  the  morphologies   of  18
high-redshift  (0.3$<z<$1.4)  LIRGs  in  the  UDF down  to  a rest-frame B-band surface
brightness of $\mu_{B0}=24.5$.  They are
a representative sample of the  LIRGs that dominate the star formation
density  at z  $\sim$ 1.  We compare  their morphologies  to  those of
MIPS-non-detected  galaxies  at similar redshift and  of  49  local  LIRGs.  We  used  the
concentration and  asymmetry indices  to quantify the  morphologies of
all these galaxies. Our main results are:

(1) The fraction of LIRGs  for $M_{B}<$-18.5 galaxies at this redshift
range  is   around  15\%,  much   higher  than  the 0.5\%  in   the  local
universe. Their morphologies are  dominated by late-type galaxies with
asymmetric (A$>$0.2) structures and  include a significant number that
appear to be mergers (40$\pm$20\%).   The morphologies of LIRGs in the
UDF  are   similar  to  those   of  local  LIRGs  but  possibly somewhat more
asymmetric. Optically luminous  MIPS-non-detected galaxies at z$\sim$1
are nearly as  asymmetric as LIRGs, in contrast  to the local universe
where  infrared-inactive galaxies  are far  more symmetric  than local
LIRGs.

(2) All of the high-z galaxy  types are relatively asymmetric, more so
than  even local  irregular  types.  Their  concentration indices  are
similiar to those of local late-type disk galaxies.  We argue that the
asymmetric  LIRGs and asymmetric  MIPS-non-detected galaxies  are from
the same parent population but  in different stages of star formation,
which suggests  the star  formation is episodic. Under this hypothesis, the fraction  of the
duration of the IR-active phase for asymmetric (A$>$0.2) galaxies is $\sim$ 25\%.

(3) The  morphologies  of local  LIRGs  and  LIRGs  at z~$\sim$~1  are
similar, suggesting that similar  conditions within the galaxy lead to
LIRG-levels  of star  formation.  At z~$\sim$~1,  such conditions  are
common  and LIRGs can  develop by  relatively minor  mergers. Locally,
most galaxies are highly symmetric, and only exceptional circumstances
such as  an interaction  with another massive galaxy  can create the  level of
asymmetry associated with LIRG-levels of star formation.

\section{Acknowledgements}
We thank the anonymous referee for very detailed and constructive comments. We also thank
John Moustakas, Jennifer Lotz, Lei Bai, Amy Stutz, Linhua Jiang and Jun Cui for helpful discussions.
This study depended on the publicly released products from the HST Ultradeep Field.
Support for this work was provided by NASA through contract 1255094 issued by JPL/California Institute of Technology.


\clearpage

\begin{landscape}

\begin{deluxetable}{llllllllllllllllllllll}
\tabletypesize{\scriptsize}
\tablecolumns{7}
\tablecaption{LIRGs in the UDF with unambiguous optical counterpart }
\tablewidth{0pt}
\tablehead{
\colhead{MIPS}     & \colhead{RA}            & \colhead{DEC}     & \colhead{ACS}        & \colhead{Sep($"$)}    & 
\colhead{redshift} & \colhead{$f_{24}$(mJy)} & \colhead{$m_{z}$} & \colhead{A}          & \colhead{C}           & 
\colhead{Type}     & \colhead{X-ray}         & \colhead{ z Source }  \\
\colhead{(1)}     & \colhead{(2)}        &   \colhead{(3)}             &  \colhead{(4)}          &
\colhead{(5)}     & \colhead{(6)}        &   \colhead{(7)}             &  \colhead{(8)}          &
\colhead{(9)}     & \colhead{(10)}       &   \colhead{(11)}            &  \colhead{(12)}         &
\colhead{(13)}                  
}
\startdata

    4641  &   03 32 35.963  &   -27 48 50.33  &    566.  &  0.16  & 1.31       & 0.06$\pm$0.012  &  21.56  & 0.21  & 3.54  &        Intermediate  &   1.31;XID-100  &  Spec  \\
    4644  &   03 32 37.512  &   -27 48 38.67  &    706.  &  0.41  & 0.67       & 0.11$\pm$0.013  &  20.87  & 0.47  & 2.51  &    Late/Interacting  &                 &  Spec   \\
    4645  &   03 32 37.172  &   -27 48 33.74  &    944.  &  0.21  & 0.85       & 0.17$\pm$0.015  &  23.57  & 0.49  & 3.19  &    Late/Interacting  &                 &  Spec   \\
    4649  &   03 32 41.078  &   -27 48 53.07  &    391.  &  0.09  & 0.68       & 0.14$\pm$0.012  &  20.21  & 0.31  & 3.41  &                Late  &                 &  Spec   \\
    5088  &   03 32 35.564  &   -27 46 27.35  &   5628.  &  1.32  & 1.08       & 0.20$\pm$0.018  &  21.12  & 0.49  & 2.46  &    Late/Interacting  &                 &36689;21.7 \\
    5092  &   03 32 34.863  &   -27 46 41.52  &   6139.  &  1.08  & 1.10       & 0.17$\pm$0.016  &  22.73  & 0.41  & 2.91  &    Late/Interacting  &                 &  Spec  \\
    5096  &   03 32 38.491  &    -27 47 2.39  &   4286.  &  0.09  & 0.92/0.97  & 0.07$\pm$0.011  &  21.19  & 0.27  & 3.80  &        Intermediate  &                 &  This work/35195,23.0    \\
    5104  &   03 32 39.865  &   -27 47 15.04  &   3739.  &  0.25  & 1.10       & 0.16$\pm$0.013  &  21.47  & 0.27  & 2.35  &                Late  &                 &  Spec   \\
    5106  &   03 32 36.931  &   -27 47 26.74  &   3285.  &  0.57  & 1.34/0.98  & 0.14$\pm$0.012  &  23.24  & 0.41  & 4.15  &   Early/Interacting  &                 &  This work/34336,23.68    \\
    5114  &   03 32 38.767  &   -27 47 32.24  &   3062.  &  0.16  & 0.46       & 0.76$\pm$0.027  &  20.88  & 0.27  & 3.06  &                Late  &   0.46;XID-567  &  Spec   \\
    5115  &   03 32 37.618  &   -27 47 43.43  &   2505.  &  0.74  & 1.17/1.06  & 0.09$\pm$0.012  &  22.11  & 0.29  & 3.36  &                Late  &                 &  This work/33813,22.85    \\
    5117  &   03 32 39.075  &    -27 46 1.84  &   6645.  &  0.12  & 1.22       & 0.26$\pm$0.016  &  20.97  & 0.07  & 2.90  &                Late  &    1.22;XID-28  &  Spec   \\
    5121  &   03 32 40.777  &   -27 46 16.01  &   5300.  &  0.32  & 0.55       & 0.14$\pm$0.013  &  20.15  & 0.27  & 2.44  &                Late  &                 &  This work    \\
    5628  &   03 32 43.227  &   -27 47 56.48  &   1802.  &  0.54  & 0.67       & 0.16$\pm$0.015  &  20.68  & 0.20  & 3.14  &                Late  &                 &  Spec  \\
    5630  &   03 32 42.275  &   -27 47 45.99  &   2344.  &  0.16  & 1.00       & 0.22$\pm$0.015  &  21.25  & 0.42  & 1.79  &    Late/Interacting  &                 &  Spec   \\
    5768  &   03 32 31.419  &   -27 47 25.04  &   3246.  &  0.82  & 0.67       & 0.12$\pm$0.010  &  21.02  & 0.46  & 2.72  &    Late/Interacting  &                 &  Spec   \\
    5828  &   03 32 39.217  &   -27 45 32.16  &   6930.  &  0.63  & 1.04       & 0.16$\pm$0.013  &  21.84  & 0.47  & 3.09  &    Late/Interacting  &                 &  Spec   \\
   13901  &   03 32 44.198  &   -27 47 32.70  &   2874.  &  0.86  & 0.74       & 0.24$\pm$0.020  &  21.42  & 0.31  & 3.24  &                Late  &                 &  Spec   \\
   15236  &   03 32 36.533  &   -27 46 30.66  &   5761.  &  1.69  & 0.71       & 0.28$\pm$0.015  &  21.21  & 0.58  & 2.54  &    Late/Interacting  &   0.77;XID-511  &36462;21.4 \\
   15942  &   03 32 44.851  &   -27 47 27.62  &   2686.  &  0.15  & 0.44       & 0.77$\pm$0.044  &  18.42  & 0.25  & 3.11  &                Late  &                 &  Spec   \\
   15943  &   03 32 45.103  &   -27 47 24.24  &   3382.  &  0.24  & 0.44       & 0.51$\pm$0.042  &  20.41  & 0.38  & 2.99  &    Late/Interacting  &   0.44;XID-646  &  Spec   \\

\enddata
\tablecomments{ Column (1): The  MIPS  ID of  the LIRGs  in  the UDF. Column(2): RA of MIPS source. Column(3): DEC of MIPS source. 
Column(4): ACS ID of opitcal counterpart of MIPS source. Column(5): The separation in arcsec between ACS source and MIPS source. Column(6): The redshift. Column(7): 24 $\mu$m flux. 
Column(8): $z$-band magnitude. Column(9): Asymmetry. Column(10): Concentration. Column(11): The galaxy type of MIPS source based on $CA$ system. Column 12: The
photometric redshift  and the catalog number of  X-ray counterparts of
MIPS  sources based  on  the  CDFS 1  MS  catalog by  \citet{Zheng04}.
Column  13:  Where  redshifts  are from  COMBO-17,  the  corresponding
catalog number along  with the magnitude at $R$  band is listed; $'$This
work$'$ indicates  that the  redshift is computed  based on  our photo-z
code \citep{Perez-Gonzalez05}. $'$Spec$'$ means the spectroscopic redshift from  VLT/FORS2 spectroscopy 
catalog \citep{Vanzella05} and VIMOS VLT Deep Survey (VVDS) v1.0 catalog \citep{LeFevre04}. }
\end{deluxetable}

\begin{deluxetable}{ccccccc}
\tabletypesize{\scriptsize}
\tablecolumns{6}
\tablecaption{The statistics of morphologies of $M_{B}<-18.5$ galaxies in the UDF and local LIRGs}
\tablewidth{0pt}
\tablehead{
\colhead{} & \multicolumn{3}{c}{  0.3$<$z$<$0.9 } &  \multicolumn{2}{c}{ 0.9$<$z$<$1.4} \\
\cline{2-4} & \cline{4-5} \\
\colhead{ }                     & \colhead{LIRGs}     & \colhead{non-LIRGs} &
\colhead{Non-detected sample}   & \colhead{LIRGs}   &
\colhead{Non-detected sample}    &   \colhead{Local LIRGs}
}
\startdata

Total number       & 10		        &             5	      &           49	  &           8       &	           88	   &           49   \\	   
Late-type          & 90$\pm$30\%    &  80$\pm$40\%   &78$\pm$13\% & 75$\pm$31\% &  81$\pm$ 10\% & 89$\pm$14\%  \\
Merging (A$>$0.35) & 40$\pm$20\%    &  20$\pm$20\%   &16$\pm$ 6\% & 63$\pm$28\% &  10$\pm$ 3\% & 31$\pm$ 8\%  \\
Symmetric (A$<$0.2)& 10$\pm$10\%    &  60$\pm$35\%   &47$\pm$ 10\% & 13$\pm$13\% &  65$\pm$ 9\% & 43$\pm$ 9\%  \\
A$^{1}$            & 0.34$\pm$0.03  &  0.22$\pm$0.05 &0.22$\pm$0.02& 0.38$\pm$0.03 &	 0.18$\pm$0.02& 0.26$\pm$0.03\\  
C$^{1}$            & 2.98$\pm$0.10  &  3.09$\pm$0.11 & 2.98$\pm$0.08 & 2.99$\pm$0.28 &  2.88$\pm$0.05  & 2.74$\pm$0.08\\  

\enddata
\tablecomments{$^{1}$Averages with standard errors of the mean. At the 0.3$<$z$<$0.9, both LIRG and MIPS-non-detected 
samples are complete. At the 0.9$<$z$<$1.4, both sample are incomplete.}
\end{deluxetable}

\clearpage
\end{landscape}

\clearpage

\end{document}